\documentstyle[aps,prl,twocolumn,graphicx]{revtex}

\newcommand{\CGO}{$\rm CuGeO_3\,$}

\def\be{\begin{equation}}
\def\ee{\end{equation}}
\def\bea{\begin{eqnarray}}
\def\eea{\end{eqnarray}}
\def\vecS{{\bf S}}

\def\Tsp{{T_{\rm\tiny SP}}}
\def\Tspo{{T_{\rm\tiny SP}^0}}
\def\xsp{{x_{\rm\tiny SP}}}
\def\calC{\mbox{$\cal{C}$}}

\begin{document}


\title{Pressure dependence and non-universal effects of microscopic
couplings on the spin-Peierls transition in CuGeO$_{\mathbf 3}$}

\author{R. Raupach, A. Kl\"umper, F. Sch\"onfeld}
\address{Institut f\"ur Theoretische Physik, Universit\"at zu K\"oln,
Z\"ulpicher Str. 77, 50937 K\"oln, Germany}

\address{~\vspace{0.5cm}\\ \parbox{14cm}{\rm The theory by Cross and
    Fisher (CF) is by now commonly accepted for the description of the
    spin-Peierls transition within an adiabatic approach. 
    The dimerization susceptibility as the essential quantity, however, 
    is approximated by means of a continuum description.
    Several important experimental observations can not
    be understood within this scope.
    Using density matrix renormalization group (DMRG) techniques we
    are able to treat the spin system exactly up to numerical
    inaccuracies. Thus we find the correct dependence of the
    equation of state on the spin-spin interaction constant $J$, 
    still in an adiabatic approach. We focus on the pressure 
    dependence of the critical temperature which is absent in the CF 
    theory as the only energy scale with considerable pressure dependence 
    is $J$ which drops out completely.
    Comparing  the theoretical findings to the experimentally measured
    pressure dependence of the spin-Peierls temperature we obtain information 
    on the variation of the frustration parameter with pressure.
    Furthermore, the ratio of the spectral gap and the transition temperature 
    is analyzed.}}

\maketitle


\section{Introduction}
Low dimensional quantum systems are currently of considerable interest
mainly due to the fascinating phase transitions driven by strong
quantum fluctuations. 
The continuous interest from the
theoretical side is provoked by the discovery of many experimental 
systems realizing quasi one-dimensional quantum systems. 
In the field of spin-Peierls systems
the discovery of the inorganic compound \CGO realized a milestone as
many measurements have been performed with high accuracy since.
Therefore, \CGO has attracted much attention in experimental as well
as in theoretical works.  The high temperature behaviour of \CGO was
found to be modelled adequately by one-dimensional frustrated
Heisenberg chains \cite{Castilla95,Riera95,Fab98,krs98}. In the dimerized phase, many
features were shown to be consistent within an adiabatic description
of the phonon degrees of freedom. This observation comprises zero
temperature \cite{Bouzerar97,lorenz98b,uhrig98a,uhrig99a}
as well as thermodynamic properties \cite{krs98}.

Even in one space dimension only a few exact results
exist particularly concerning thermodynamics. For integrable systems
the thermodynamical potentials and asymptotic behaviour of correlation
functions are known. A notorious problem is posed by response functions
and non-integrable systems in general. With respect to this, the
recently developed transfer matrix DMRG (TMRG)
\cite{Bursill96,Wang97,Shibata97} on the basis of transfer matrices
\cite{SuzukiI87} provides a very powerful method to calculate
thermodynamic quantities of spin chains without any use of
perturbative methods. This has been demonstrated in several
applications \cite{krs98,Shibata97,Maisinger98,Shibata98,Coombes98,Xiang98,Eggert98}.
 
In this paper we study the influence of microscopic coupling constants
on the spin-Peierls transition temperature. This allows for an
understanding of the considerable pressure dependence of the phase
diagram along the following line of reasoning. It is known that
external pressure affects the magnetic properties of \CGO considerably 
\cite{Fab98,Takahashi95,loosdrecht97}.
Fits for the magnetic susceptiblity yield the change of the
nearest-neighbour spin interaction $J$ and estimates for the
frustration paramter $\alpha$.  Using these data
for $J$ as function of pressure we are able to explain the observed
increase of the spin-Peierls temperature and estimate the
pressure dependence of the next-nearest-neighbour exchange

The outline of the paper is as follows. In section \ref{model} we
present the model and a motivation for our description of the
experimentally studied spin-Peierls systems. 
We study the static dimerization susceptibility in section \ref{response}.
Section \ref{pressure} is
devoted to the computation of the critical temperature as function of
the spin exchange couplings and external pressure, respectively. We
give a comparison of our results with experimental measurements. In
section \ref{BCS} we investigate
the spectral gap and its ratio to the spin-Peierls temperature.
The conclusion is given in section \ref{conclusion}.

\section{Model}
\label{model}
In the inorganic spin-Peierls compound \CGO the magnetic interactions
are attributed to Heisenberg spin exchange. There is numerous evidence
that in addition to the nearest-neighbour interaction ($J$) a
next-nearest-neighbour exchange $J' = \alpha J$
\cite{Castilla95,Riera95,Fab98} with $\alpha=0.35$ has to be taken
into account.  Usually the constant $\alpha$ is referred to as
frustration parameter.

At the spin-Peierls transition the system undergoes a structural
phase transition driven by the quantum spin system coupled to the 
phonons. The spin-phonon coupling is modelled by
spin exchange integrals depending linearily on the local displacements.
The adiabatic treatment
yields a quantum spin system coupled to just one phonon mode
with the commonly used Hamiltonian
\begin{equation}
  \hat{H} = \sum_i \left\{ J \left[(1+\delta_i)\vecS_i\vecS_{i+1} + 
\alpha\vecS_i\vecS_{i+2}\right] + \frac{K}{2}\delta_i^2\right\},
\label{Ham}
\end{equation}
where $\vecS_i$ are spin 1/2 operators, $\delta_i=(-1)^i \delta$
denotes the modulation of the magnetic exchange couplings in the
dimerized phase. Here, we restrict ourselves to vanishing external
magnetic field where the system shows a phase transition from
the uniform (U), i.e. $\delta=0$ to the dimerized (D) phase $(\delta >
0)$.

The elastic energy can be expressed in terms of microscopic constants
rendering (\ref{Ham}) equivalent to an RPA treatment of the phonon
propagator for the full spin-phonon system. Within RPA the
condition for the phase transition is identical to that for
(\ref{Ham}) as formulated in (\ref{phasetrans}) below if $K$ is
adjusted to the following value \cite{werner98}
\begin{equation}
 \label{K_J}
 K = \frac{J^2}{2} \left(\sum_i\frac{\lambda_i^2}{M_i \Omega_i^2}\right)^{-1} =: \mbox{$\cal{C}$} J^2.
\end{equation}
The sum runs over the spin-Peierls active modes. For each mode
$i$, $M_i$ denotes the effective mass of the unit cell, $\Omega_i$ the
frequency of the spin-Peierls active phonon and $\lambda_i$ the
spin-phonon coupling constant. In particular, $K$ is proportional to
$J^2$. The constant $\cal{C}$ contains only the microscopic
parameters of the underlying lattice. 

Our numerical investigations do not improve over those of CF \cite{CF79} with
respect to the RPA treatment. However, within the RPA approximation we
deal with the complete dynamics of the quantum spin system.  Note that
in Ref. \cite{CF79} a continuum description of the spin system was used with
a subsequent bosonization treatment which is believed to capture only the
long distance asymptotics of the correlation functions.

\section{Response functions}
\label{response}
The static dimerization susceptibility of the spin
system is defined by
\begin{equation}
 A_{\alpha}(x) = - J^{-1} \lim_{\delta\to 0} \frac{\partial^2 f_{\alpha}(x, \delta)}
{\partial \delta^2},
\end{equation}
where $x=T/J$, and $f_{\alpha}(x,\delta)$ is the free energy per site for
system (\ref{Ham}) with fixed dimerization $\delta$, frustration $\alpha$ and $K$ set to zero. The response function $A$ is nothing but the correlation of 
the nearest neighbour spin
exchange $\vecS_i \vecS_{i+1}$ (dimer operator) at momentum $q=\pi$
and energy $\omega=0$. The U/D phase transition takes place for
\begin{equation}
 \label{phasetrans} A_{\alpha}(\xsp) = K/J = \mbox{$\cal{C}$} J.
\end{equation}
For details the reader is referred to \cite{krs98}. 

Let us now review the results obtained by CF.
Within the bosonization approach they find $x \cdot A_{\alpha=0}(x) = \chi_0$
with $\chi_0 \approx 0.26$. As a direct consequence by use of
the inversion of (\ref{phasetrans}), 
\begin{equation}
  \label{Tsp_von_J}
  \Tsp = J A_{\alpha}^{-1}(\mbox{$\cal{C}$} J),
\end{equation}
this yields $d \Tsp / d J = 0$. Of course, this is also
clear from the fact that the energy scale of the spin system
completely drops out due to scale invariance.

\begin{figure}
\center{\includegraphics[width=\columnwidth]{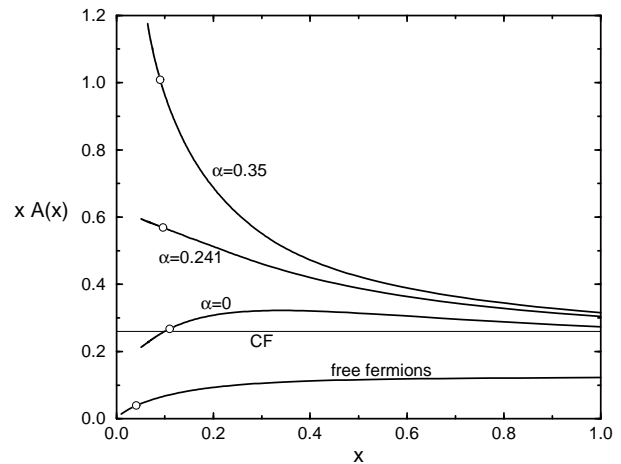}}
\caption[T_mal_AT] {Depiction of the function $x\cdot A(x)$ with $x=T/J$
for: free fermions, Heisenberg model in continuum limit (CF),
and TMRG results for frustration parameter
$\alpha=0,0.241,0.35$. The circels denote the relevant values for \CGO
(see text).}
\label{T_mal_AT}
\end{figure}

Fig. \ref{T_mal_AT} shows a comparison of our TMRG
results\footnote{We have used $24$ states in the renormalization step
  and have set the accuracy with respect to Trotter decomposition to
  $(TM)^{-1}=0.05$.} for the function $x \cdot A_{\alpha}(x)$ for various
values of $\alpha$ with the findings of CF
and exact data for free fermions. The enormous progress achieved by the
numerical analysis is the correct treatment of the spin system on the
lattice at practically all length scales. This improves over the
continuum limit approach in which the asymptotics of correlation
functions is incorrectly extended to short distances.
For the unfrustrated Heisenberg model, i.e.  $\alpha=0$, we are
able to observe directly the deviations between the continuum limit
and a lattice treatment of the spin degrees of freedom.

With respect to \CGO we fix $J$ by the requirement that the
experimental magnetic susceptibility equals that of the strictly
one-dimensional model \underline{at} the critical point leading to
$J=130K~(\alpha=0)$, $J=150K~(\alpha\approx\alpha_c\approx 0.2412$ \cite{Castilla95,Eggert96,Okamoto93}), $J=160K~(\alpha=0.35)$, and
$J=350K$ (free fermions). The circles in Fig.~\ref{T_mal_AT} denote the
values of $x\cdot A_{\alpha}(x)$ at the experimentally determined spin-Peierls temperature $\Tspo=14.4K$ for \CGO (see e.g.~\cite{hirota94,lorenz97a,pouget94}). These values imply constants (cf.~Eq.~\ref{phasetrans}) $\calC_{0}\approx0.019~\mbox{K}^{-1}~(\alpha=0)$, $\calC_{0.241}\approx0.040~\mbox{K}^{-1}~(\alpha=0.241)$, $\calC_{0.35}\approx0.070~\mbox{K}^{-1}~(\alpha=0.35)$, and $\calC_{\mbox{\tiny{ff}}}\approx0.0027~\mbox{K}^{-1}$ (free fermions).

For the unfrustrated model the value $\xsp\cdot A_{\alpha=0}(\xsp)$ 
almost coincides with the results of CF. However the agreement happens only 
fortuitously since $\chi_0$ is a zero temperature quantity.
For other frustration parameters qualitative and
quantitative deviations from the CF-line appear. The divergence of
$A(x)$ for free fermions, Heisenberg model with $\alpha < \alpha_c$, $\alpha=\alpha_c$,
and $\alpha > \alpha_c$, is $\log(x)$, 
$1/x \times \mbox{log. corrections}$\cite{krs98,Kluemper99},
$1/x$ (see \cite{Eggert96} and references therein), and exponential \cite{krs98}, respectively. The
quantitative values for $\xsp \cdot A(\xsp)$ also differ from each other. We
must conclude that it is risky to deduce quantitative results from the
bosonization approach as already pointed out by CF. 
For applications to \CGO it is furthermore uncertain if the
pure Heisenberg chain can even yield the qualitative results
correctly.  From the analysis of the susceptibility data at higher
temperature we are led to favour the frustration parameter
$\alpha=0.35$ \cite{Riera95,Fab98,krs98} for which the behaviour of the response function 
deviates considerably from $\chi_0/x$.

\section{Pressure Dependence of the Magnetic System}
\label{pressure}
From (\ref{Tsp_von_J}) it is straight forward to deduce the dependence of the spin-Peierls
temperature on the variation of the spin coupling constants. The relation is valid for every fixed value of $\alpha$. We first focus on the system with $\alpha_0=0.35$ at ambient pressure. 

\begin{figure}
\center{\includegraphics[width=\columnwidth]{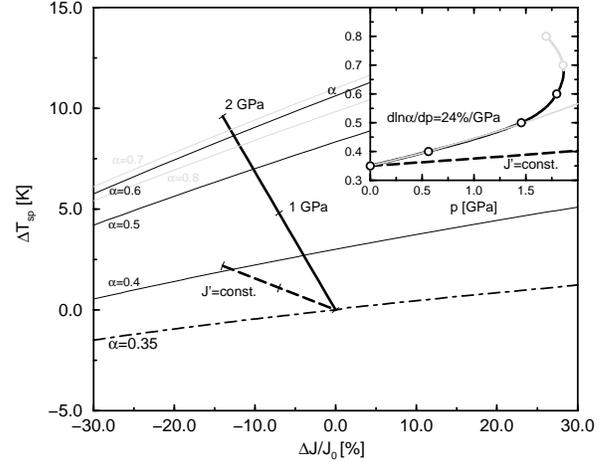}}
\caption[Tsp_J_0.35] {Variation of the spin-Peierls temperature versus the relative change of the magnetic exchange coupling. The lines show TMRG results for constant $\alpha$ and fixed $\calC_{0.35}$. The thick black line displays the experimental behaviour as an implicit function of pressure. The dashed (dashed-dotted) line shows the theoretical results for $dJ^{\prime} / dp = 0$ ($d\alpha / dp = 0$). The corresponding dependencies of $\alpha$ on the pressure are shown in the inset.}
\label{Tsp_J_0.35}
\end{figure}

The dependence of $J$ and $\alpha$ on the external hydrostatic pressure have already been obtained from the relation between magnetostriction and the pressure dependence of the magnetic susceptibility $\chi$ \cite{Fab98}. From our TMRG data we find $\frac{\partial \chi}{\partial \ln J} \gg \frac{\partial \chi}{\partial \ln \alpha}$ at constant temperature. These two quantities appear in the following expression, $\frac{d \chi}{d p} = \sum_i \frac{\partial \chi}{\partial \ln x_i} \frac{\partial \ln x_i}{\partial p}$, where $x_1=J$ and $x_2=\alpha$.
Therefore, we conclude the value for the pressure dependence of $J$ to be more reliable than that of $\alpha$. The authors of Ref. \cite{Fab98} deduced a value of $\frac{d \ln J}{d p}=-7.0(5)\%/{\rm GPa}$ using the relation between magnetostriction and pressure dependencies of $J$ and $\alpha$. 
Using in addition the experimentally determined value for the pressure dependence of the spin-Peierls temperature $\frac{d \Tsp}{dp} \approx 4.8$~K/GPa \cite{Takahashi95} we obtain the relation $\Tsp(J)$ (thick black line in Fig.~\ref{Tsp_J_0.35}).

Due to the particular geometry involved in the super-exchange mechanism we expect the magnetic exchange energies to respond much more sensitively to the pressure (bond-bending mechanism \cite{khomskii96,khomskii97}) than the phonon frequencies or the spin-phonon coupling constants. We therefore consider $\calC$ as independent of pressure which applicability for \CGO will break down at higher pressure. The numerical results with $\calC=\calC_{0.35}$ are shown in Fig.~\ref{Tsp_J_0.35}. Obviously, a constant value of $\alpha=0.35$ (dashed-dotted line) can not explain the observed behaviour, even yielding a change of the critical temperature to opposite direction. 

From geometrical reasons we expect $J^{\prime}$ to be non-decreasing. Assuming $J^{\prime} =\mbox{constant}$ to be realized, the increase of $\Tsp$ is too small by a factor of approximately 4 (as displayed by the thick dashed line in Fig.~\ref{Tsp_J_0.35}). Consequentially, the main effect must be a strong increase of $J^{\prime}$. The deduced dependence of $\alpha$ on the hydrostatic pressure fits well with $\frac{d \ln \alpha}{d p} = 24\pm 2\%/{\rm GPa}$ (or $\frac{d \ln J^{\prime}}{d p} = 17 \pm 2\%/{\rm GPa}$ respectively) up to about 1.4~GPa as displayed in the inset of Fig.~\ref{Tsp_J_0.35}. The error was determined assuming that the value of $\frac{d \Tsp}{dp}$ involves an error of $\pm 0.5K/\mbox{GPa}$, where the uncertainty in $\frac{d \ln J}{d p}$ only playes a secondary role. In general, already the weak requirement that $J$ is non-increasing with hydrostatic pressure gives a minimum pressure dependence of $\frac{d \ln \alpha}{d p} \ge 20\%/{\rm GPa}$. The pressure dependence of $\alpha$ has already been investigated in Ref.~\cite{Fab98}, however, with large error bounds which are respected by our results. The divergence of $\frac{d \ln \alpha}{d p}$ at a finite pressure or in other words, an upper limit of the critical temperature, is a physical prediction of the chosen one-dimensional approach. It is mostly based on a limited spontaneous gap as a function of $\alpha$. But as already mentioned above, we expect an agreement with \CGO only in the low pressure region.

A similiar analysis can be done on the basis of the unfrustrated Heisenberg model and free fermions. Evaluating the magnetic susceptibility data of Takahashi et al. \cite{Takahashi95} we derive $\frac{d \ln J}{d p} \approx -5\%/{\rm GPa}$ ($\alpha=0$) and $\frac{d \ln J}{d p} \approx -6\%/{\rm GPa}$ (free fermions). 
In contrast to the frustrated case, a change of $\alpha$ under pressure for
these initially unfrustrated models is not reasonable. 
The deduced theoretical BCS-ratio is too small by a factor of about 3 for free fermions (or equivalently Hartree-Fock calculations), in the case of the unfrustrated Heisenberg model even by a factor of $\approx 15$ (see Tab.~I).  

We must conclude that neither the unfrustrated Heisenberg model nor Hartree-Fock results are able to describe the physics of \CGO correctly. 
In constrast to this a Heisenberg model with parameters $\alpha_0=0.35$ ans $\frac{d \ln \alpha}{d p} = 24\%/{\rm GPa}$ turns out to reproduce the experimental findings up to about 1.4~GPa.

\section{``BCS-ratio''}
\label{BCS}
\begin{figure}
\center{\includegraphics[width=\columnwidth]{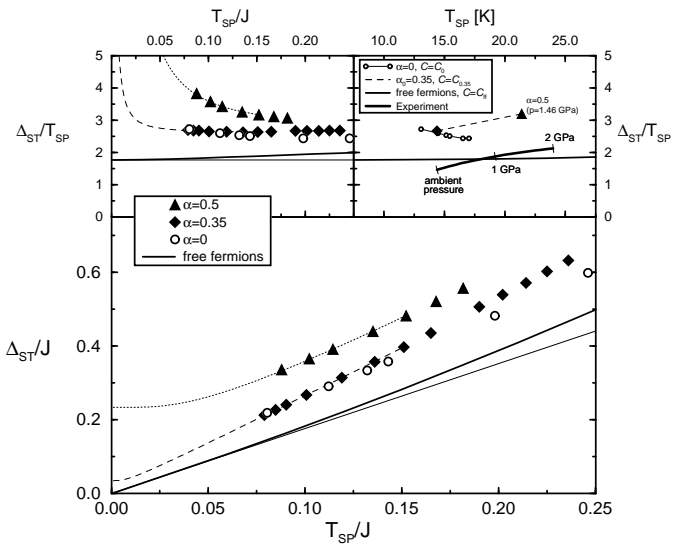}}
\caption[BCS_ratio] {Bottom: Singlet-triplet gap $\Delta_{\mbox{\tiny ST}}$ as 
   a function of the spin-Peierls temperature. 
   Symbols denote the DMRG results, the
   solid line shows the exact result for free fermions. The thin lines
   show the expected behaviour from the scaling of $A$ (see text). Top left:
   Ratio of the spin gap and spin-Peierls temperature (BCS ratio). Top
   right: BCS ratio, experiment versus theory.}
\label{BCS_ratio}
\end{figure}

Another interesting quantity is the ratio of the singlet-triplet gap to the spin-Peierls temperature, known as ``BCS-ratio''. Combining the TMRG data for $A$ with zero temperature DRMG calculations for the singlet-triplet gap as a function of $\alpha$ and the dimerization one gets a relation between $\Delta_{\mbox{\tiny ST}}$ and $\Tsp$. The results are shown in the bottom section of Fig.~\ref{BCS_ratio}.  

Using the scaling of $A$ \cite{krs98}, the definition of the critical temperature  and the dependence of the ground state energy on small saturation dimerizations $\delta_0 = \delta(T=0)$ one finds 
\begin{equation}
 \delta_0 \simeq
    {\Tsp}^{3/2} \exp(-\frac{\Delta_{\mbox{\tiny ST}}^0}{2}/\Tsp), 
\end{equation}
ignoring logarithmic corrections for $\alpha < \alpha_c$.
Here $\Delta_{\mbox{\tiny ST}}^0$ denotes the singlet-triplet gap for vanishing dimerization, which is zero for $\alpha \le \alpha_c$. In the next step we apply $\Delta_{\mbox{\tiny ST}} - \Delta_{\mbox{\tiny ST}}^0 \simeq \delta_0^{2/3}$ \cite{Affleck89,Uhrig99} to derive an explicit asymptotic relation between the gap and the spin-Peierls temperature,
\begin{equation}
 \label{Delta_scaling}
 \Delta_{\mbox{\tiny ST}} - \Delta_{\mbox{\tiny ST}}^0 \simeq 
    \Tsp \exp(-\frac{\Delta_{\mbox{\tiny ST}}^0}{3}/\Tsp),
\end{equation}
again neglecting logarithmic corrections for $\alpha < \alpha_c$.
The thin lines in Fig.~\ref{BCS_ratio} show fits according to Eq.~(\ref{Delta_scaling}) in the range up to $T/J=0.15$. For $\alpha=0.5$ we used the known value $\Delta_{\mbox{\tiny ST}}^0 = 0.2338J$ \cite{caspe84,Mue-Ha98}, hence only performed a one-parameter fit in the presented region. For $\alpha=0.35$ a spontanous gap of $\Delta_{\mbox{\tiny ST}}^0 = 0.035J$ is used, also derived by $T=0$ DMRG. For $\alpha=0$ the predicted linear behaviour (cf.~(\ref{Delta_scaling})) can not be seen due to the logarithmic corrections. A linear extrapolation of our data points shows a positive offset. However, this is consistent with the presence of logarithmic corrections since we expect a zero limit at $T=0$ with infinite slope. The free fermions show the well known behaviour $\Delta_{\mbox{\tiny ST}} / \Tsp \approx 1.76$. We now derive the BCS-ratio as a function of the spin-Peierls temperature simply by dividing by $\Tsp$ (upper left of Fig.~\ref{BCS_ratio}). 
Fistly we observe that the Heisenberg like models have a distinctly larger BCS-ratio than free fermions. 

For comparison to experiment we again fix the constant $\calC$. Furthermore, the pressure dependence of $J$ and the frustration derived in section \ref{pressure} is taken into account here. The results are shown in the top right of Fig.~\ref{BCS_ratio}. 
Interestingly, the experimental result \cite{Takahashi95} compares well with the low
temperature asymptotics for free fermions but showing a pronounced increase with pressure which is not reproduced. From the investigations in Ref. \cite{krs98} we already know for Heisenberg models that the gap and therefore the BCS-ratio is larger than the experimentally observed one. This happens due to the strictly one-dimensional treatment of \CGO, i.e. the neglect of the dispersion perpendicular to the chain which will lower the true gap. The unfrustrated chain even yields a qualitatively incorrect tendency of a decreasing BCS-ratio under pressure the appropriately frustrated system is at least able to explain the increase. 

\section{conclusion}
\label{conclusion}
The TMRG analysis of the spin-Peierls phase transition allows a complete treatment of the quantum dynamics. In contrast to continuum descriptions correlations are respected at all length scales, which leads to the exact response functions. Within the scope of an adiabatic description, equivalent to RPA as in the CF theory, we are able to study the influence of the spin-spin exchange energy scale $J$ on the critical temperature. We like to emphasize that these results are a non-trivial improvement over the CF theory which shows no dependence on $J$ at all. Moreover, frustrated models can be investigated as well.

Using the pressure dependence of $J$ it is possible to study the dependence of the spin-Peierls temperature on pressure. Neither Hartree-Fock calculations nor the unfrustrated Heisenberg chain yield the strong increase as measured. Once again, our investigations favour a frustration of $\alpha=0.35$ for \CGO at ambient pressure. We find a rather strong dependence of the frustration on pressure, $\frac{d \ln \alpha}{d p} = 24 \pm 2 \%/{\rm GPa}$, which agrees with earlier studies \cite{Fab98}.

The analysis of the ``BCS-ratio'' also gives a clear indication that frustration is present in \CGO, even though some quantitative deviations can only be explained by residual perpendicular couplings. There is evidence for a strong dependence of $\alpha$ on pressure from the ``BCS-ratio''.

\begin{center}
{\bf Acknowledgement}
\end{center}
We like to thank U.~L\"ow for carefully reading the manuscript.
The work was supported by the Deutsche Forschungsgemeinschaft through SFB 341.

\begin{table}
\begin{tabular}{c|c|c|c}
 & $\frac{d \ln J}{dp} [\%/\mbox{GPa}]$ & $\frac{d \ln \alpha}{dp} [\%/\mbox{GPa}]$ & $\frac{d \Tsp}{dp} [\mbox{K/GPa}]$ \\ \hline
free fermions & $\approx -6$ & (0) & $\mathbf{\approx}${\bf 1.8} \\ \hline
$\alpha = 0$ & $\approx -5$ & (0) & $\mathbf{\approx}${\bf 0.3} \\ \hline
$\alpha_0 = 0.35$ & $-7.0\pm 0.4$ & {\bf 24}$\pm${\bf 2} & $4.8 \pm 0.5$\\ 
\end{tabular}
\caption{Pressure dependences of the spin system parameters and the spin-Peierls temperature as explained in the text. Numbers in bold face denote results obtained by our DMRG calculations, all other quantities were used as input data. Note that models based on free fermions or unfrustrated Heisenberg chains can not explain the rather large pressure dependence of the critical temperature.}
\end{table}

\end{document}